\documentclass[aps,prb,twocolumn,groupedaddress]{revtex4}

\usepackage{amsmath}
\usepackage{epsfig}
\usepackage{graphics}

\def\ba{\begin{array}}
\def\ea{\end{array}}

\begin{document}


\title{Cyclostationary measurement of low-frequency odd moments of current fluctuations}


\author{Tero T. Heikkil\"a}
\email[Correspondence to ]{Tero.T.Heikkila@hut.fi}

\author{Leif Roschier}

\affiliation{Low Temperature Laboratory, P.O. Box 2200, FIN-02015
HUT, Finland}


\date{\today}

\begin{abstract}
Measurement of odd moments of current fluctuations is difficult
due to strict requirements for band-pass filtering. We propose how
these requirements can be overcome using cyclostationary driving
of the measured signal and indicate how the measurement accuracy
can be tested through the phase dependence of the moments of the
fluctuations. We consider two schemes, the mixing scheme and the
statistics scheme, where the current statistics can be accessed.
We also address the limitations of the schemes, due to excess
noise and due to the effects of the environment, and, finally,
discuss the required measurement times for typical setups.
\end{abstract}


\maketitle


\section{Introduction}
\label{sect:intro}

The discrete nature of the charge carriers shows up in the
peculiar statistics of the transmitted current through mesoscopic
systems.\cite{levitovetal} Consequently, these statistics in
general cannot be described through the usual Gaussian probability
density which is completely determined by its first two moments
(cumulants), average current and noise, whereas its higher
cumulants vanish (for a difference between the moments and
cumulants, see Ref.~\onlinecite{weissteinworldofmath}). Recently,
there have been many theoretical predictions for the behavior of
either the third moment of fluctuations or the full couting
statistics in different types of mesoscopic systems (see a review
in Ref.~\onlinecite{naznoisebook}), including normal-metal tunnel
junctions, \cite{levitovreznikov} normal-metal--superconductor
\cite{muzykantskii94,belzig01b} and superconductor--superconductor
junctions, \cite{belzig01a,cuevas03,johansson03} diffusive
\cite{lee96,nagaev02,pilgramup04,gutmangefen} and chaotic wires,
\cite{pilgram03} double-barrier junctions \cite{dejong96} and
Coulomb-blockaded systems. \cite{bagrets03} For example, the
second and third moments of current fluctuations through a
phase-coherent scatterer at $T=0$ may be expressed as
\begin{equation}
\begin{split}
S\equiv \langle \delta I \delta I \rangle &= \frac{e^3 |V|}{h}
\sum_n T_n (1-T_n) \equiv F_2 e |I|\\
R \equiv \langle \delta I \delta I \delta I \rangle &= \frac{e^4
V}{h} \sum_n T_n (1-T_n) (1-2 T_n) \equiv F_3 e^2 I,
\end{split}
\end{equation}
where $T_n$ are transmission eigenvalues. Their distribution
depends on the properties of the scatterer, \cite{beenakkerrmp}
and it can be characterized by defining the Fano factors $F_2$ and
$F_3$. Compared to the average current $I$, the higher moments may
thus reveal additional information about the scatterer (for an
example, see Ref.~\onlinecite{reuletup04}). Moreover, $R \neq 0$
implies that the distribution of fluctuating currents is "skew".
This has important consequences in the situations where the driven
current fluctuations act as an environment to another mesoscopic
system. \cite{sonin04}

\begin{figure}[h]
\centering \includegraphics[width=\columnwidth]{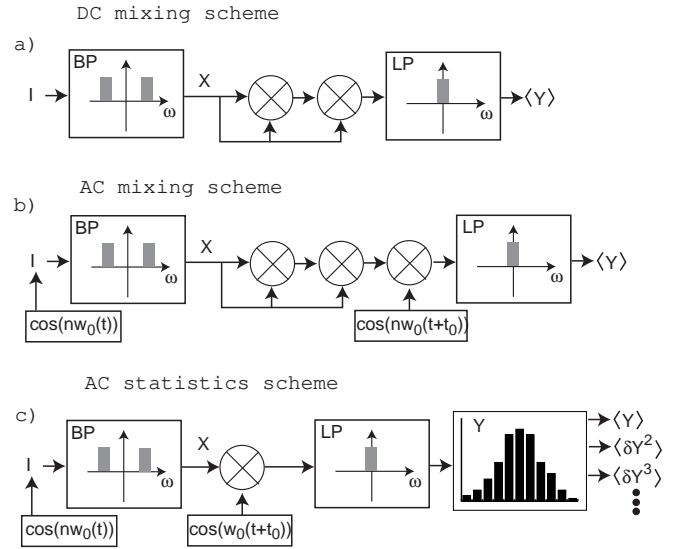}
\caption{Measurement schemes considered in this paper. Each
contain an input $I$ filtered through a band-pass filter (denoted
with BP, corresponding to the function $H_{BP}$). In the "DC
mixing scheme", analogous to that utilized in
Ref.~\onlinecite{reulet03}, the stationarily driven and band-pass
filtered input is mixed twice with itself, low-pass filtered
(denoted with $LP$, corresponding to the function $H_{LP}^m$) and
averaged. The "AC mixing scheme" is a modification of the DC
scheme, including a cyclostationary driving of the measured
system, and mixing with an extra function
$f(t)=\cos(n\omega_0(t+t_0))$, synchronized with the driving and
with a controlled phase shift $t_0$. In the "AC statistics
scheme", the statistics of the output is measured and the driven
and filtered signal is mixed only with $f(t)$.}
\label{fig:schemes}
\end{figure}

However, up to date there exists only one published measurement of
the higher (than second) moments. \cite{reulet03,reuletup04} One
of the reasons for this is the necessary conditions for the
filtering in these devices: Impedance matching between the
amplifier and the sample is possible typically only within a
fairly narrow frequency band, outside of which the signal from the
sample does not couple to the measurement device. However, as we
show below, for the typical measurements of the odd $n$'th moments
of fluctuations, the requirement for the bandwidth $2\delta
\omega$ of the measured frequencies around the mean frequency
$\omega_0$ is $\delta \omega
> \omega_0/n$. \cite{reulet03} Especially for the measurement of
high-impedance samples ($Z$ larger than 1 k$\Omega$) this
requirement is very hard if not impossible. In this paper, we show
how this requirement can be circumvented by a cyclostationary
driving of the measured system (see Fig.~\ref{fig:schemes}).

To motivate the use of cyclostationary driving, consider the
$n$'th central moment of current fluctuations in the frequency
space 
\begin{equation}
\tilde{M}^{(n)}(\omega_1, \omega_2, \dots, \omega_n) \equiv
\langle \delta I(\omega_1) \delta I(\omega_2) \dots \delta
I(\omega_n) \rangle,
\end{equation}
where the brackets denote ensemble averaging and $\delta I(\omega)
\equiv I(\omega)-\langle I(\omega) \rangle$. If the studied system
is time independent, i.e., if it is driven with a constant
voltage, we get the signal only from the frequencies that satisfy
\begin{equation}
\omega_1 + \omega_2 + \dots \omega_n=0. \label{eq:wcondition}
\end{equation}
Consider now a situation where the signal is band-pass filtered
before its correlators are measured, such that only the
frequencies in the band $\omega \in [\omega_0-\delta \omega,
\omega_0 + \delta \omega]$ pass through the filters. If $\delta
\omega \ll \omega_0$, we cannot fulfill condition
(\ref{eq:wcondition}) for odd $n$. Simple algebra shows that for
the $n$'th moment (where $n$ is odd), we only get some signal if
\begin{equation}
\delta \omega > \omega_0/n.
\end{equation}
However, this condition is true for {\it stationary} signal, but
if we adiabatically drive the system with frequency $\omega_D$,
condition (\ref{eq:wcondition}) changes to
\begin{equation}
\omega_1' + \omega_1 + \dots \omega_{n-1}=m\omega_D,
\end{equation}
with an integer $m$. Reasonable examples considering the odd
$n$'th moments are $\omega_D=\omega_0, 3 \omega_0, \dots, n
\omega_0$.

In the second section of this paper, we detail two specific
measurement schemes using cyclostationary driving and compare them
to that applied in Ref.~\onlinecite{reulet03}. The first of the
schemes is based on mixing the signal with itself and with an
oscillatory function $f(t)$, and the second on measuring the
statistics of the output signal from the sample after filtering
and mixing with $f(t)$. The third section discusses the effect of
cyclostationary driving in some example systems and the fourth
considers the limitations of the presented measurement schemes due
to the excess noise in the environment, and estimates the required
averaging time for obtaining an accurate signal. The details of
the calculations are presented in the Appendix.

\section{Measurement schemes}
\label{sect:schemes}

Let us consider the specific measurement schemes for measuring the
moments of current fluctuations depicted in
Fig.~\ref{fig:schemes}. The emphasis of this analysis is on the
third moment, but the results and the schemes are fairly
straightforward to generalize also to higher moments. In both of
these schemes, the studied system is driven with a time-dependent
voltage of the form (for most of the discussion below, the
different harmonics do not need to be in the same phase, but we
assume them in phase for simplicity)
\begin{equation}
V(t)=V_0 + V_1 \cos(\omega_0 t) +  V_3 \cos(3 \omega_0 t).
\label{eq:acvoltage}
\end{equation}
This voltage produces in general an output current of the form
\begin{equation}
I(t)=\sum_l I_l \cos(l\omega_0 t). \label{eq:accurrent}
\end{equation}
In linear systems with resistance $R$, $I_l=V_l/R$ and thus
$I_l=0$ for $l>3$. In nonlinear systems, the output current would
also contain higher harmonics of $\omega_0$, but these do not
contribute essentially to the results. However, the coefficients
$I_i$ should be taken from the proper Fourier analysis of the
output current.

With pure voltage driving, the moments of the low-frequency
current fluctuations are proportional to either the average
current $I(t)$ (for odd moments) or its absolute value (for even
moments). This holds as long as the driving is adiabatic, i.e.,
the highest driving frequency $3\omega_0$ is low compared to the
internal energy scales of the probed system or to $eV(t)/\hbar$.
\cite{pilgramup04,galaktionovup03} In fact, any slow measurement
of the signal as a function of the driving voltage can be
considered as "adiabatic" in this sense --- the only difference to
that considered here is that typically the time scales of such
voltage variations are of the order of seconds, much slower than
the scales for band-pass filtering. Furthermore, for low
frequencies, the fluctuations can be considered "white", i.e.,
they are of the form (below, we reserve the symbol $S$ for the
second moment, i.e., noise, and use the symbol $R$ for the third
moment)\cite{numbertwo}
\begin{equation}
S(t,t') \equiv \langle \delta I(t) \delta I(t') \rangle = \sum_{l}
S_l \cos(l\omega_0 t) \delta(t-t') \label{eq:whitenoise}
\end{equation}
and
\begin{equation}
\begin{split}
R(t,t',t'') &\equiv \langle \delta I(t) \delta I(t') \delta I(t'')
\rangle \\&=  \sum_{l} R_l \cos(l\omega_0 t)
\delta(t-t')\delta(t-t''). \label{eq:acthird}
\end{split}
\end{equation}
Similarly, the $n$'th central moment would be
\begin{equation}
\begin{split}
&M^{(n)}(t_1,t_2,\dots,t_n) \equiv \langle \delta I(t_1) \delta
I(t_2) \dots \delta I(t_n) \rangle \\ &= \sum_{l} M_l^{(n)}
\cos(l\omega_0 t_1) \delta(t_1-t_2) \delta(t_1-t_3) \dots
\delta(t_1-t_n).\label{eq:acnthmom}
\end{split}
\end{equation}
In most cases below, it is enough to cut the series in the $n$'th
harmonic (for the $n$'th moment) as the higher ones do not
contribute to the results. There are many sources for the higher
harmonics than those present in the driving and one has to always
first do the proper Fourier analysis to find these, prior to
examining the results: i) finite temperature of the order of the
voltage $V$ makes the relations $S(V)$ and $R(V)$ nonlinear even
in many linear systems, \cite{blanterbuettiker,gutmangefen} ii) if
the sign of the total current is allowed to vary, one has to take
the Fourier components from the absolute value of the current for
$S(V)$, and iii) these harmonics arise naturally in nonlinear
systems. Examples of these are discussed in
Sect.~\ref{sect:examples}. However, we stress that all these
effects can only make rise to frequency components of the form
$l\omega_0$, where $l$ is some integer.

For what follows, the assumption of the "white" fluctuations is in
fact less strict than assumed in
Eqs.~(\ref{eq:whitenoise}-\ref{eq:acnthmom}): it is enough that
the fluctuations are frequency independent in the window allowed
by the band-pass filter.

For both schemes, the signal is assumed to be filtered with an
ideal band-pass filter allowing frequencies $\omega \in
[\omega_0-\delta \omega, \omega_0+\delta \omega]$. Such a filter
can be described through the boxcar function
$\tilde{H}_{BP}(\omega) \equiv \theta(|\omega|-(\omega_0-\delta
\omega)) - \theta(|\omega| - (\omega_0 + \delta \omega))$, where
$\theta(\omega)$ is the Heaviside step function. Thus, the
outgoing signal from the filter is in the time domain a
convolution (below, unless it is obvious, the functions in the
frequency domain are denoted with a tilde)
\begin{equation}
X(t)=\frac{1}{\sqrt{2\pi}} H_{BP}(t) \ast I(t).
\end{equation}
At the end of the processing, the signal is gathered for a time
exceeding $1/\omega_0$. Such a process can be described by a
convolution with a low-pass filter $\tilde{H}_{LP}^m(\omega) =
1-\theta(|\omega|-\omega_m)$, where $\omega_m \ll \omega_0$.

\subsection{Mixing schemes}

In the mixing schemes depicted in Fig.~\ref{fig:schemes} (schemes
a and b), we assume that the signal is, after filtering, twice
mixed with itself and with a function $f(t)$ specified below,
low-pass filtered, and finally ensemble averaged. Thus, the
measurement result is
\begin{equation}
\langle Y \rangle \equiv \frac{1}{\sqrt{2\pi}} \left\langle
H_{LP}^m \ast \left[ f(t) X^3(t)  \right] \right\rangle.
\end{equation}
Assuming that over a sequence of measurements performed for the
averaging, the filters and the function $f(t)$ do not essentially
change, one obtains
\begin{widetext}
\begin{equation}
\begin{split}
\langle Y \rangle &= \frac{1}{4\pi^2} H_{LP}^m \ast \left[f(t')
\langle (H_{BP} \ast I)^3 \rangle\right]
\\&= \frac{1}{4\pi^2} H_{LP}^m \ast \left[f(t') \int dt_1 dt_2 dt_3 H_{BP}(t'-t_1)H_{BP}(t'-t_2)H_{BP}(t'-t_3)
\langle I(t_1) I(t_2) I(t_3) \rangle\right].
\end{split}
\end{equation}
\end{widetext}
Below, we take a closer look at two ways of driving the signal:
first with a DC bias and wide band-pass filters, and then with an
AC bias and narrow band-pass filters.

\subsubsection{DC mixing scheme}

Assume first that the system is DC biased, i.e., only the first
term in Eqs.~(\ref{eq:acvoltage}-\ref{eq:acthird}) is nonzero. In
this case, we do not need the function generator, i.e., $f(t)=1$.
Such a setup corresponds to the one utilized in
Ref.~\onlinecite{reulet03} and it is depicted in
Fig.~\ref{fig:schemes}a.

One can write the current in the form $I(t)=\langle I \rangle +
\delta I(t)$, where $\delta I(t)$ is a zero-mean fluctuating
current. Then the third raw moment of current fluctuations is
\begin{equation}
\begin{split}
&\langle I(t) I(t') I(t'') \rangle = \langle I \rangle^3 + \\&
\langle I \rangle S (\delta(t-t') + \delta(t-t'') +
\delta(t'-t'')) + R \delta(t-t') \delta(t-t'').
\end{split}
\end{equation}
The average signal coming out of the sampling filter $H_{LP}^m$ is
\begin{widetext}
\begin{equation}
\begin{split}
&\langle Y \rangle = \frac{1}{4\pi^2} \langle H_{LP}^m \ast (H_{BP} \ast I(t))^3 \rangle \\
&= \frac{1}{4\pi^2}H_{LP}^m \ast \int dt_1'' dt_2'' dt_3''
H_{BP}(t'-t_1'') H_{BP}(t'-t_2'')
H_{BP}(t'-t_3'') \langle I(t_1'') I(t_2'') I(t_3'') \rangle \\
&= \frac{1}{4\pi^2} H_{LP}^m \ast \left[\left(\langle I \rangle
\int dt'' H_{BP}(t'-t'')\right)^3 + 3 \langle I \rangle S
\left(\int dt'' H_{BP}(t'-t'')\right)\left(\int dt''
H_{BP}^2(t'-t'')\right) + R \int dt'' H_{BP}^3(t'-t'')\right].
\end{split}
\end{equation}
\end{widetext}
The band-pass filter filters out the DC signal,
\begin{equation}
\int dt'' H_{BP}(t-t'') = \sqrt{2\pi} \tilde{H}_{BP}(0) = 0
\end{equation}
as long as it is a band-pass filter, i.e., $\delta \omega <
\omega_0$. Therefore, only the last term remains, and 
we obtain
\begin{equation}
\langle Y \rangle 
= R \frac{3}{4\pi^2} (3\delta \omega - \omega_0)^2 \theta(3\delta
\omega - \omega_0).
\end{equation}
It can be seen that for $\delta \omega < \omega_0/3$, $\langle Y
\rangle = 0$, in agreement with the discussion in
Sect.~\ref{sect:intro}. The scaling of the observed quantity as a
function of the bandwidth was applied in
Ref.~\onlinecite{reulet03} to show that it was indeed the third
moment that was measured.

\subsubsection{AC mixing scheme}
\label{subs:acmixing}
Assume that the current, noise and the third
moment are of the form given in Eqs.~(\ref{eq:accurrent}),
(\ref{eq:whitenoise}), and (\ref{eq:acthird}), respectively. For
simplicity, we assume that the lowest driving frequency equals the
center frequency of the band-pass filter and denote both by
$\omega_0$. The average output signal is given by
\begin{widetext}
\begin{equation}
\begin{split}
\langle Y \rangle=&\frac{1}{4\pi^2} H_{LP}^m \ast \bigg\{ f(t')
\bigg[\left(\sum_{l} I_l \int dt'' H_{BP}(t'-t'') \cos(l\omega_0
t'')\right)^3 \\&+ 3\left(\sum_l I_l \int dt'' H_{BP}(t'-t'')
\cos(l\omega_0 t'')\right)\left(\sum_l S_l \int dt''
H_{BP}^2(t'-t'') \cos(l\omega_0 t'')\right) \\&+ \sum_l R_l \int
dt'' H_{BP}^3(t'-t'')\cos(l\omega_0 t'')\bigg]\bigg\}.
\end{split}
\end{equation}
Now we have three types of integrals with the band-pass filter
$H_{BP}$. 
We obtain
\begin{align}
\int dt'' H_{BP}(t'-t'') \cos(l\omega_0 t'') &= \sqrt{2\pi}
\tilde{H}_{BP}(l\omega_0) \cos(l\omega_0 t') = \sqrt{2\pi}
\delta_{l,1}
\cos(\omega_0 t')\\
 \int dt'' H_{BP}^2(t'-t'') \cos(l\omega_0 t'') &=
(\tilde{H}_{BP} \ast \tilde{H}_{BP})|_{\omega=l\omega_0}
\cos(l\omega_0 t') = 2\delta\omega\left[2\delta_{l,0} +
\delta_{l,2}
\cos(2\omega_0 t')\right],\\
\int dt'' H_{BP}^3(t'-t'') \cos(l\omega_0 t'') &=
\frac{1}{\sqrt{2\pi}}(\tilde{H}_{BP} \ast \tilde{H}_{BP} \ast
\tilde{H}_{BP})|_{\omega=l\omega_0} \cos(l\omega_0 t') =
\frac{3\delta\omega^2}{\sqrt{2\pi}}\left[3\delta_{l,1}
\cos(\omega_0 t') + \delta_{l,3}\cos(3\omega_0 t')\right].
\end{align}
Combining these results, we get
\begin{equation}
\langle Y \rangle = \frac{H_{LP}^m}{\sqrt{8\pi^3}} \ast \left\{
f(t')\left[2\pi I_1^3 \cos^3(\omega_0 t') + 6I_1 \delta \omega
\left[2S_0\cos(\omega_0 t') + S_{2}\cos(2\omega_0
t')\right]+\frac{3\delta\omega^2}{2\pi}\left[3R_1 \cos(\omega_0
t') + R_{3} \cos(3\omega_0 t')\right]\right]\right\}.
\end{equation}
\end{widetext}
Now it is time to specify the mixing function $f(t)$. This mixes
some of the above oscillating functions to the zero frequency and
thereafter the low-pass filter $H_{LP}^m$ filters the other
frequencies out. There are two meaningful choices for $f(t)$, (a)
$f(t)=\cos(\omega_0 (t+t_0))$ and (b) $f(t)=\cos(3\omega_0
(t+t_0))$, where $t_0$ represents the phase shift between the
function generator and the signal. With the first choice, the
result is
\begin{equation}
\langle Y_a \rangle = \left[\frac{3}{8} I_1^3 + 6\frac{\delta
\omega}{2\pi} I_1 S_0 + \frac{9\delta\omega^2}{8\pi^2} R_1\right]
\cos(\omega_0 t_0),
\end{equation}
and with the second,
\begin{equation}
\langle Y_b \rangle = \left(\frac{1}{8} I_1^3 + \frac{3\delta
\omega^2}{8\pi^2} R_{3}\right) \cos(3\omega_0 t_0).
\label{eq:mixingresult}
\end{equation}
This result illustrates that, in a linear system, setting
$I_0=I_1=I_2=0$, the signal with $f(t)=\cos(3\omega_0 (t+t_0))$ is
dependent only on the third moment $R_3$. Moreover, the dependence
on the phase shift shows that the function $f(t)$ and the driving
have to be synchronized. This phase dependence can also be
utilized as a signature of the fact that the measured quantity
really corresponds to the properties of the signal coming from the
sample, analogously to the dependence on the bandwidths utilized
in Ref.~\onlinecite{reulet03}.

\subsection{AC statistics scheme}
\label{subs:acstatistics} In the second scheme
(Fig.~\ref{fig:schemes}c), the AC-driven current is mixed only
once with the function $f(t)=\cos(\omega_0 (t+t_0))$ and the
output signal
\begin{equation}
Y=\frac{1}{2\pi} H_{LP}^m \ast \left[f (H_{BP} \ast I)\right]
\end{equation}
is measured many times. This produces an ensemble of $Y$-values,
whose moments are related to the moments of the current $I(t)$. As
above, we assume that the bandwidths of both the band-pass filter,
$\delta \omega$, and of the sampling, $\omega_m$, are much lower
than $\omega_0$.

The $n$'th central moment of this $Y$-distribution is given by
\begin{widetext}
\begin{equation}
\langle (\delta Y)^n \rangle = \sum_{l=0}^\infty
\frac{1}{(2\pi)^n} \int \left[\prod_{i=1}^n dt_i H_{LP}^m(t_i)
f(t_i)\right] \int \left[\prod_{i=1}^n dt_i'
H_{BP}(t_i-t_i')\right] \langle \prod_{i=1}^n \delta I(t_i')
\rangle_l,
\end{equation}
where $l$ labels harmonics of the base frequency $\omega_0$.
Utilizing the white-noise assumption as in
Eqs.~(\ref{eq:whitenoise}-\ref{eq:acnthmom}) yields
\begin{equation}
\langle (\delta Y)^n \rangle = \sum_{l=0}^\infty
\frac{1}{(2\pi)^n} \int \left[\prod_{i=1}^n dt_i H_{LP}^m(t_i)
f(t_i)\right] \int dt' \left[\prod_{i=1}^{n} H_{BP}(t_i-t')\right]
M_l^{(n)} \cos(l\omega_0 t'),
\end{equation}
where $M_l^{(n)}$ is the coefficient of the $l$'th harmonic of the
$n$'th central moment in the measured signal. This integral is
evaluated in Appendix \ref{sect:secondschemeapp}. The results for
the six lowest moments are (the first, average, is the raw moment,
the rest are central moments, i.e., defined w.r.t. the average)
\begin{align}
\langle Y \rangle &= \frac{1}{2} I_1 \cos(\omega_0 t_0) \label{eq:firstm}\\
\langle (\delta Y)^2 \rangle &= \frac{\omega_d}{4\pi} \left[2S_0 +
S_2
\cos(2\omega_0 t_0)\right]\label{eq:secondm}\\
\langle (\delta Y)^3 \rangle &= \frac{3\omega_d^2}{32\pi^2}
\left[3R_1
\cos(\omega_0 t_0) + R_3 \cos(3\omega_0 t_0)\right]\label{eq:thirdm}\\
\langle (\delta Y)^4 \rangle &= \frac{\omega_d^3}{24\pi^3}\left[6
M_0^{(4)} + 4 M_2^{(4)} \cos(2\omega_0 t_0) + M_4^{(4)}
\cos(4\omega_0 t_0)\right]\label{eq:fourthm}\\
\langle (\delta Y)^5 \rangle &= \frac{115 \omega_d^4}{6144 \pi^4}
\left[10M_1^{(5)} \cos(\omega_0 t_0) + 5 M_3^{(5)} \cos(3\omega_0
t_0)
+ M_5^{(5)} \cos(5\omega_0 t_0)\right]\label{eq:fifthm}\\
\langle (\delta Y)^6 \rangle &= \frac{11\omega_d^5}{1280
\pi^5}\left[20 M_0^{(6)} + 15 M_2^{(6)} \cos(2\omega_0 t_0) + 6
M_4^{(6)} \cos(4\omega_0 t_0) + M_6^{(6)} \cos(6\omega_0
t_0)\right],\label{eq:sixthm}
\end{align}
\end{widetext}
where $\omega_d=\min(\omega_m,\delta \omega)$. Note that which
harmonics are nonzero depends on how the system is driven. For
example, if one drives the system with a voltage $V_3
\cos(3\omega_0 t)$, the only finite moments (among the first six)
contributing to the measured signal are $S_0$, $R_3$, $M_0^{(4)}$,
$M_3^{(5)}$, $M_0^{(6)}$ and $M_6^{(6)}$. Such examples are
discussed in more detail in the following section.

\section{Examples of higher harmonics}
\label{sect:examples} In this section, we discuss examples of
systems where one could measure the third moment of fluctuations
with AC driving. The considered examples are the simplest linear
systems at low temperatures, the finite-temperature case where
part of the signal is mixed to higher harmonics, and thirdly a
generic example of a nonlinear system.

\subsection{Linear systems at low temperatures}
\label{subs:linearsyst} The simplest but yet important example of
a system where our theory is applicable is a linear system with
resistance $R$, driven with an AC voltage $V(t)=I_3 R
\cos(3\omega_0 t)$. In a tunnel junction, neglecting the effects
of its environment, \cite{reulet03,beenakker03} the third moment
is independent of the temperature \cite{levitovreznikov} whereas
the noise is in general a mixture of shot noise (produced by
driving) and thermal noise. \cite{blanterbuettiker} In other
systems, also the third moment becomes temperature dependent.
\cite{gutmangefen} For $eV \gg k_B T$, the output current, noise
and the third moment are
\begin{align}
I(t)&=I_3 \cos(3\omega_0 t)\\
S(t)&=eF_2 I_3 |\cos(3\omega_0 t)| = eF_2 I_3 (\frac{2}{\pi} +
\frac{4}{3\pi} \cos(6\omega_0 t) + \dots)\\
R(t)&=e^2F_3 I_3 \cos(3\omega_0 t).
\end{align}
Here $F_2$ and $F_3$ are the Fano factors ($F_2=F_3=1$ for a
tunnel junction) for the second and third moments, respectively.
Applying the AC mixing scheme and mixing with $f(t)=\cos(3\omega_0
(t+t_0))$ yields the output
\begin{equation}
\langle Y \rangle = \frac{3}{2}\left(\frac{\delta
\omega}{2\pi}\right)^2 e^2 F_3 I_3.
\end{equation}
In the AC statistics scheme, the average of the output signal
vanishes, the variance is
\begin{equation}
\langle (\delta Y)^2 \rangle = \frac{\min(\omega_m,\delta
\omega)}{\pi^2} eF_2 I_3 \label{eq:varlinTzero}
\end{equation}
and the third moment is
\begin{equation}
\langle (\delta Y)^3 \rangle =
\frac{3}{8}\left(\frac{\min(\omega_m,\delta
\omega)}{2\pi}\right)^2 \cos(3\omega_0 t_0) e^2 F_3 I_3.
\label{eq:thirdlinTzero}
\end{equation}
This illustrates that, as long as the thermal noise and
environmental noise are negligible, the best signal from the third
moment (compared to the second) is obtained with wide-band
filtering but low-amplitude driving.

For a specific example, assume a realistic set of numbers,
$I_3=100$ nA and $\min(\omega_m,\delta \omega)/(2\pi) = 10$ MHz.
Then, we get
\begin{align}
\langle Y^2 \rangle & \approx F_2 ({\rm 320 pA})^2\label{eq:varex}\\
\langle Y^3 \rangle &\approx F_3 (46 {\rm pA})^3\cos(3\omega_0
t_0)\label{eq:thirdex}
\end{align}
Thus, the third moment should be well observable in this
situation. With larger currents or smaller bandwidths, the second
moment becomes relatively larger, but only as $(\langle Y^2
\rangle)^{1/2}/(\langle Y^3 \rangle)^{1/3} \propto (I_3/(e\delta
\omega))^{1/6}$.

\subsection{Diffusive wire at a finite temperature}
\label{subs:finitetemp} The temperature dependence of the higher
moments of fluctuations depends on the studied system. For
example, for a diffusive wire one expects \cite{gutmangefen}
\begin{align}
S&=\frac{eI}{3}\coth\left(\frac{eV}{2k_B T}\right)+\frac{4k_B T}{3 Z_s}\\
R&=e^2 I \frac{6(-1+e^{4p})+(1-26e^{2p}+e^{4p})p}{15
p(-1+e^{2p})^2}.
\end{align}
Here $p=eV/2k_B T$. In this case, for an oscillating voltage, some
of the moments mix into higher frequencies, altering the expected
outcome of the measurement. However, in the extremal cases $eV_i
\ll k_B T$ and $eV_i \gg k_B T$, where $eV_i$ is the amplitude of
the oscillating voltage, this mixing is not important and one may
simply use
\begin{align}
S(eV_i \ll k_B T)&=\frac{2 k_B T}{Z_s}\\
S(eV_i \gg k_B T)&=\frac{1}{3}e|I_i(t)|\\
R(eV_i \ll k_B T)&=\frac{1}{3}e^2 V_i(t)\\
R(eV_i \gg k_B T)&=\frac{1}{15}e^2 V_i(t).
\end{align}
Here $Z_s$ is the resistance of the measured sample.

For example, let us assume that we drive the system with the
voltage
\begin{equation}
V(t)=V_3 \cos(3\omega_0 t).
\end{equation}
As a diffusive wire is a linear system, the average current
follows the oscillations of the voltage, $I(t)=I_3 \cos(3\omega_0
t)$. Our measuring schemes yield information about the stationary
part $S_0$ of the noise, and about the $3\omega_0$-component $R_3$
of the third moment. These are plotted in
Fig.~{\ref{fig:noiseandthirdwithtemp} as a function of $eV_3/2k_B
T$. One can observe that in this particular case, the crossover to
the "pure shot noise" takes place for $eV_3 \gtrsim 10 k_B T$
whereas the crossover between the low- and high-temperature third
moments is wider, saturating only at some $eV_3 \gtrsim 40 k_B T$.

\begin{figure}[h]
\centering
\includegraphics[width=\columnwidth]{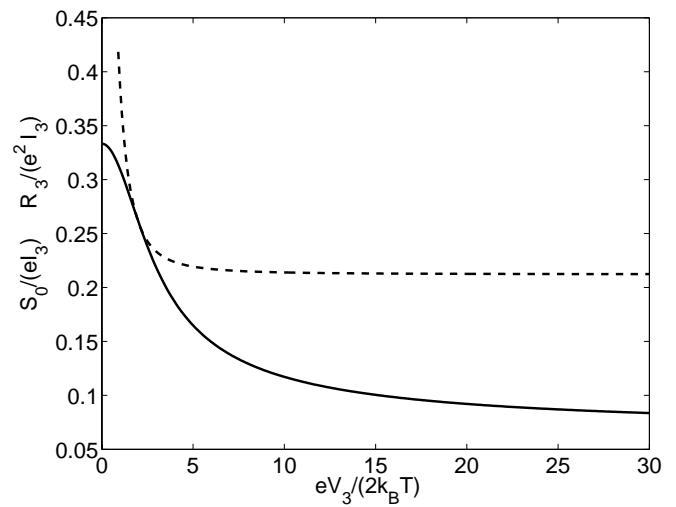}
\caption{The stationary part $S_0-4k_B T/(3Z_s)$ of the second
moment (dashed line) and the lowest harmonic $R_3$ of the third
moment (solid line) in a diffusive wire driven with $V_3
\cos(3\omega_0 t)$, normalized to $eI_3$ and $e^2 I_3$,
respectively. Note that the previous diverges with $eV_3 \ll k_B
T$ as then $S_0$ contains only the thermal noise and is
essentially independent of $I_3$.}
\label{fig:noiseandthirdwithtemp}
\end{figure}

\subsection{Generic nonlinear system}
\label{subs:genericnonlinear}
 Many interesting mesoscopic systems
are inherently nonlinear: there exists an energy scale $E_c$ below
and above which the electron transport mechanism may differ. Such
a behavior is often signalled by different Fano factors for the
second and third moments below and above $E_c$. Below, we explain
how the moments can be measured in such systems applying
cyclostationary driving. There are two extreme alternatives:
either one drives the system with a single oscillating voltage,
say, $V(t)=V_3 \cos(3\omega_0 t)$, or with a large DC voltage
$V_0$ fixing the operating point and with a small AC excitation of
the form $V_3 \cos(3\omega_0 t)$ on top of it. The first
alternative has the benefit of minimizing the excess noise
compared to the third moment (important for the averaging time),
whereas in the latter case the results are easier to interpret.

In nonlinear systems, the nonlinearities mix the driving
frequencies to higher harmonics of the base frequency. To see an
example of this, we consider a generic non-linear system at zero
temperature, assuming that the differential conductance of the
system is
\begin{equation}
\frac{dI}{dV} = \begin{cases} G_a, \quad V < V_c\\
                              G_b, \quad V > V_c.
                              \end{cases}
\end{equation}
Here $V_c$ is some characteristic voltage describing the
nonlinearity. For example, in the case of Coulomb blockade, it
would be determined by the charging energy, or in the case of a
normal-metal - insulator - superconductor (NIS) junction, by the
superconducting gap. Typically in such systems, the process for
charge transport below and above $V_c$ differs (for example, in
the case of intermediate-transparency NIS junction, for $V <
\Delta/e$, it would be due to Andreev reflection, and for $V >
\Delta/e$, due to quasiparticles). Therefore, also the
differential Fano factor may differ \cite{stenberg02} between the
two regimes. Thus, let us assume that the differential noise
(second moment) is given by
\begin{equation}
\frac{dS}{dV} = \begin{cases} e F_2^a G_a {\rm sgn}(V), \quad V < V_c\\
                              e F_2^b G_b {\rm sgn}(V), \quad V > V_c.
                              \end{cases}
\end{equation}
Similarly, the third moment would be
\begin{equation}
\frac{dR}{dV} = \begin{cases} e^2 F_3^a G_a, \quad V < V_c\\
                              e^2 F_3^b G_b, \quad V > V_c.
                              \end{cases}
\end{equation}
Given a slowly varying voltage $V(t)$, the resulting second and
third moments would be
\begin{align}
S(t)&=\begin{cases} e F_2^a G_a |V(t)|, \quad |V(t)| < V_c\\
                   e |F_2^a G_a V_c + F_2^b G_b (V(t)-V_c)|, \quad
                   |V(t)| > V_c
                   \end{cases}\label{eq:gennonlinsec}\\
R(t)&=\begin{cases} e^2 F_3^a G_a V(t), \quad |V(t)| < V_c\\
                   e^2 F_3^a G_a V_c + F_3^b G_b (V(t)-V_c), \quad
                   |V(t)| > V_c.
                   \end{cases}
\end{align}
Now drive the system with $V(t)=V_3 \cos(3\omega_0 t)$, $V_3 >
V_c$. The DC part of the noise, contributing to the measurement
outcome, is
\begin{equation}
S_0 = \frac{2eV_3}{\pi}\left[F_2^a G_a + (F_2^b G_b - F_2^a G_a)
  \sqrt{1-\left(\frac{V_c}{V_3}\right)^2}\right].
\end{equation}
The lowest harmonic of the third moment, oscillating with the
frequency $3\omega_0 t$ would then be
\begin{equation}
\begin{split}
R_3 &= e^2 F_3^a G_a V_3 + \frac{2e^2}{\pi} (F_3^b G_b - F_3^a
G_a)
\\ &\times \left(V_3 \arccos\left(\frac{V_C}{V_3}\right)+V_C
\sqrt{1-\left(\frac{V_C}{V_3}\right)^2}\right).
\end{split}
\end{equation}
For large-amplitude driving, $V_3 \gg V_c$, one obtains
\begin{align}
S_0 &\rightarrow \frac{2e}{\pi} F_2^b G_b V_3 + o\left(\left(\frac{V_c}{V_3}\right)^2\right)\\
R_3 &\rightarrow e^2 V_3 F_3^b G_b +
o\left(\left(\frac{V_c}{V_3}\right)^3\right).
\end{align}
Hence, as one would expect, large-amplitude driving is mostly
sensitive to the behavior at large voltages.

Another approach would be to DC bias the system to a given
operating point $V_0$, and add a non-stationary term $V_3
\cos(3\omega_0 t)$ on top of this (this type of measurement is
frequently used for the differential conductance). In this way,
the measured third moment would be sensitive to that near the
operating point only. In this case, the second moment would be of
the form of Eq.~\eqref{eq:gennonlinsec} with $V(t) \approx V_0$,
only weakly dependent on $V_3$ (except perhaps for $V_0 \approx
V_c$). On the contrary, the measured third moment would follow
$V_3$,
\begin{equation}
R_3=e^2 F_3^o G_o V_3
\end{equation}
where $F_3^o$ is the Fano factor for the third moment and $G_o$
the differential conductance at the operating point. This approach
would in principle work in any kind of nonlinear systems, provided
that $V_3$ is chosen appropriately. However, this approach has the
disadvantage that the second moment is relatively much larger than
the third, making the averaging time long (see
Sect.~\ref{sect:limitations}).

\section{Limitations of measurement}
\label{sect:limitations}

In this section, we discuss the limitations in the measurement of
higher moments of current fluctuations with the emphasis on the
third moment. We consider the effects of environmental noise added
to the signal after amplifying, the effect of a Gaussian noise in
the electromagnetic environment, and the averaging time.

\subsection{Effect of amplifier noise}

In typical measurement systems, the signal coming from the filters
has to be amplified before it is mixed and detected. This
amplification introduces additional noise $\delta I_a(t)$ in the
output signal. In the following, we analyze the effect of such
noise, assuming that it is Gaussian and uncorrelated with the
fluctuations in the current coming from the sample. The second
moment of such fluctuations is
\begin{equation}
S_a(\omega) \equiv \frac{1}{\sqrt{2\pi}} \int d(t-t') \langle
\delta I_a(t) \delta I_a(t') \rangle e^{i\omega(t-t')},
\end{equation}
whereas the average and the other odd moments vanish. Moreover,
one has to assume that this noise is intrinsically band-limited
(due to finite $RC$-times etc.), i.e.,
\begin{equation}
\langle \delta I_a(t) \delta I_a(t) \rangle = \frac{1}{\sqrt{2}}
\int d\omega S_a(\omega) \equiv \frac{1}{\sqrt{2}} S_{\rm tot}^a
\end{equation}
is finite. Due to this noise, the fluctuating signal before mixing
is of the form
\begin{equation}
X(t)= \frac{1}{\sqrt{2\pi}} H_{BP}(t) \ast I(t) + \delta I_a(t).
\end{equation}
In the AC mixing scheme, the resulting signal after mixing is
\begin{equation}
\langle Y \rangle = \langle Y \rangle_0 +
\frac{3}{2\pi}\left\{H_{LP}^m \ast \left[f(t)(H_{BP}(t) \ast
\langle I(t) \rangle) \langle \delta I_a(t)^2
\rangle\right]\right\},
\end{equation}
where the first term has been calculated above and the second is
due to excess noise. Such noise contributes only if the signal
after band-pass filtering has a finite average and we apply
$f(t)=\cos(\omega_0 (t+t_0))$,
\begin{equation}
\langle Y \rangle = \langle Y \rangle_0 + \frac{3}{4\pi} S_{\rm
tot}^a I_1 \cos(\omega_0 t_0).
\end{equation}
For $f(t)=\cos(3\omega_0 (t+t_0))$, the amplifier noise
contribution vanishes.

In the AC statistics scheme, the amplifier noise does not
contribute to the average signal. The second moment becomes
\begin{equation}
\langle (\delta Y)^2 \rangle = \langle (\delta Y)^2 \rangle_0 +
\frac{1}{\sqrt{2\pi}} S_a(\omega_0) \omega_m,
\end{equation}
where the first term is that calculated above, and the second is
due to amplifier noise within the band $\omega \in
[\omega_0-\omega_m,\omega_0+\omega_m]$. As in the AC mixing
scheme, the effect of amplifier noise is finite only for a
non-vanishing average signal,
\begin{equation}
\langle (\delta Y)^3 \rangle = \langle (\delta Y)^3 \rangle_0 +
\frac{3}{4\pi} S_a(\omega_0) \omega_m \langle I_1 \rangle
\cos(\omega_0 t_0).
\end{equation}
Apart from the noise in the amplifiers between the filtering and
mixing, there may be other sources of noise. The most important
noise source, that in the electromagnetic environment seen by the
sample is treated below. Another source of noise, possibly
relevant in the mixing schemes, is the noise added by the mixers.
However, there the dependence of the signal on the phase $t_0$ can
be used to distinguish the measured signal from the noise.

\begin{figure}[h]
\centering
\includegraphics[width=\columnwidth]{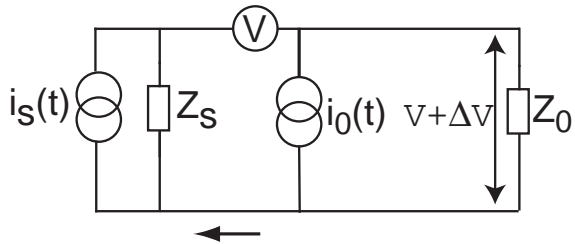}
\caption{Circuit studied for the effect of fluctuations in the
electromagnetic environment on the third moment of the measured
fluctuations. Circuit consists of the measured system with
impedance $Z_s$ and fluctuating current $i_s(t)$, coupled to its
environment with impedance $Z_0$ and Gaussian current fluctuations
$i_0(t)$. The arrow on the bottom of the figure indicates the
chosen reference direction for the current.}
\label{fig:environmenteffect}
\end{figure}

\subsection{Linear system with environment}
\label{subs:enveffect} The effect of electromagnetic environment
becomes the more important the higher moments one measures. For
the second moment, the noise in the environment simply adds to
that in the measured sample and scales the signal from the sample
through the impedance ratio. Recently, Beenakker, Kindermann and
Nazarov \cite{beenakker03} showed that the environment has a
strong effect on the third moment, as this includes the effect of
current noise in the environment biasing the sample. In the
following, we generalize their approach to the case of
nonstationary driving. We also address the methods for filtering
in these devices and show how this yields the results of the
general analysis in Sect.~\ref{sect:schemes}.

To properly deal with the filtering, let us assume that the
measured system can be separated in two parts (see
Fig.~\ref{fig:environmenteffect}): measured sample with impedance
$Z_s$ and a band-pass filter with impedance $Z_0(\omega)$. The
impedances of the measuring lines, current amplifiers, etc. can be
included in $Z_0(\omega)$. For simplicity, we assume that within
the considered frequency interval, $Z_s$ is frequency independent,
and $Z_0(\omega)$ is given by
\begin{equation}
Z_0(\omega)=\begin{cases} Z_p, \quad \omega \in \pm
[\omega_0-\delta
\omega,\omega_0 + \delta \omega]\\
            Z_e, \quad \text{otherwise}.
\end{cases}
\end{equation}
Here, $Z_e$ is either much larger (in the case when one would
measure directly the current fluctuations) or much smaller (when
converting the current fluctuations to voltage fluctuations) than
the sample impedance. For simplicity, all impedances are chosen
real. Moreover, assume that the current fluctuations in
$Z_0(\omega)$ are Gaussian with the second moment given by
\begin{equation}
S_I^f(\omega,\omega')\equiv \langle \delta I_0(\omega) \delta
I_0(\omega')\rangle =\frac{2k_B T_0}{Z_0(\omega)}\delta(\omega +
\omega').
\end{equation}
Here $T_0$ characterizes the noise temperature of the environment
and in general it may differ from the temperature of the sample.

The intrinsic current fluctuations $\delta I(\omega)$ in the two
impedances cause voltage fluctuations $\Delta V(\omega)$ over
them. As a result, the total current fluctuations at the two
impedances are of the form
\begin{align}
\Delta I_0(\omega) &= \delta I_0(\omega) - \frac{\Delta
V(\omega)}{Z_0(\omega)}\\
\Delta I_s(\omega) &= \delta I_s(\omega) + \frac{\Delta
V(\omega)}{Z_s(\omega)}.
\end{align}
Assuming low enough frequencies (compared to the frequencies
describing charge relaxation), the two currents should equal. This
requirement leads to
\begin{align}
\Delta V(\omega)&=\frac{Z_s Z_0(\omega)}{Z(\omega)}[\delta
I_0(\omega) - \delta I_s(\omega)]\\
\Delta I(\omega)&=\frac{Z_s}{Z(\omega)} \delta I_s(\omega) +
\frac{Z_0(\omega)}{Z(\omega)} \delta I_0(\omega).
\end{align}
Here we defined $Z(\omega) = Z_s + Z_0(\omega)$. The second
moments of current fluctuations in the total system and voltage
fluctuations over the impedances are given by\cite{matchingnote}
\begin{equation}
\begin{split}
S_I(\omega,\omega') &\equiv \langle \Delta I(\omega) \Delta
I(\omega')\rangle \\&= \frac{Z_s^2}{Z(\omega)Z(\omega')}
S_I^s(\omega,\omega') +
\frac{Z_0(\omega)Z_0(\omega')}{Z(\omega)Z(\omega')}
S_I^f(\omega,\omega')
\end{split}
\end{equation}
\begin{equation}
\begin{split}
S_V(\omega,\omega') &\equiv \langle \Delta V(\omega) \Delta
V(\omega')\rangle \\&= \frac{Z_s^2 Z_0(\omega)
Z_0(\omega')}{Z(\omega) Z(\omega')}(S_I^s(\omega,\omega') +
S_I^f(\omega,\omega')),
\end{split}
\end{equation}
where $S_I^s(\omega,\omega') \equiv \langle \delta I_s(\omega)
\delta I_s(\omega') \rangle$ measures the second moment of current
fluctuations in the sample.

One can easily find that if the fluctuations are measured from the
current, the optimal situation for noise measurement is $Z_p \ll
Z_s$ and $Z_e \gg Z_s$ whereas for the measurement of voltage
fluctuations, the maximum signal is obtained with $Z_p \approx
Z_s$ and $Z_e \ll Z_s$.\cite{optimumnote} In practice, when the
signal power has to be compared to the resolution of the meters,
the latter case is advantageous.

The major invention by Beenakker {\it et al.} was the fact that as
the magnitude of current fluctuations in mesoscopic samples
depends on the amplitude of driving, the voltage fluctuations in
the environment tune these fluctuations, and thereby induce
cross-correlations between the current fluctuations in the sample
and in the environment. Following their work, we expand the second
moment of fluctuations, $S_I^e(V)$ in the driving voltage near the
average voltage,
\begin{equation}
\begin{split}
&\delta I_s(\omega) \delta I_s(\omega') = \int dt
e^{i(\omega+\omega') t} (\delta I_s(t))^2
\\&\approx \int dt e^{i(\omega + \omega') t} \left(S_I^s(\langle
V(t) \rangle) + \Delta V(t) \frac{dS_I^s(V)}{dV}|_{V=\langle V(t)
\rangle}\right).
\end{split}
\end{equation}
The first-order fluctuation term is enough to describe the
environmental effect on the measured third moment. Now assume the
driving voltage has the form given by Eq.~(\ref{eq:acvoltage}) and
expand the above two terms in Fourier harmonics,
\begin{align}
S_I^s(\langle V(t) \rangle) &= \sum_{n=0}^\infty S_n
\cos(n\omega_0 t)\\
\frac{dS_I^s(V)}{dV}|_{V=\langle V(t) \rangle} &=
\sum_{n=0}^\infty D_n \cos(n\omega_0 t).
\end{align}
For example, in a linear system in the limit $eV \gg k_B T$,
driving with $V(t)=V_3 \cos(3\omega_0 t)$ yields
\begin{equation}
D_1=0, \quad D_3=\frac{4F_2e}{\pi Z_s},
\end{equation}
where $F_2$ is the Fano factor for the second moment.

\begin{widetext}
Combining, we get
\begin{equation}
\delta I_s(\omega) \delta I_s(\omega') \approx \frac{1}{2} \sum_n
\left[S_n(\delta(\omega + \omega'+n\omega_0) +
\delta(\omega+\omega'-n\omega_0)) + D_n(\Delta V(\omega + \omega'
+ n\omega_0) + \Delta V(\omega + \omega' - n\omega_0))\right].
\end{equation}
The third moment of the current fluctuations $\Delta I(\omega)$
contains three parts: first coming from the "intrinsic" third
moment of the measured sample, $R_I^s$, second from the
cross-correlations between the sample and its environment, and the
third from the current fluctuations of the sample driving the
sample itself \cite{beenakker03}
\begin{equation}
\begin{split}
&R_I(\omega,\omega',\omega'')\equiv \langle \Delta I(\omega)
\Delta I(\omega') \Delta I(\omega'') \rangle =
\left(\frac{Z_s^3}{Z(\omega) Z(\omega') Z(\omega'')}\right)\Bigg[
R_I^S(\omega,\omega',\omega'') \\&+ k_B T_0
\left(\frac{Z_0(\omega)}{Z(\omega)} +
\frac{Z_0(\omega')}{Z(\omega')} +
\frac{Z_0(\omega'')}{Z(\omega'')}\right) \sum_{n,\alpha=\pm 1} D_n
\delta(\omega_s+n\alpha\omega_0)\\ &- \frac{Z_s}{4}
\sum_{\overset{n,m}{\alpha,\beta=\pm 1}} D_n S_m
\left(\frac{Z_0(\omega+m\beta  \omega_0)}{Z(\omega+m\beta
\omega_0)}+\frac{Z_0(\omega'+m\beta  \omega_0)}{Z(\omega'+m\beta
\omega_0)}+\frac{Z_0(\omega''+m\beta  \omega_0)}{Z(\omega''+m\beta
\omega_0)}\right) \delta(\omega_s+(n \alpha + m
\beta)\omega_0)\Bigg],
\end{split}
\end{equation}
where $\omega_s=\omega+\omega'+\omega''$. This correlator is
essential if one is able to directly measure the current
fluctuations. In this case, specifying $Z_e \gg Z_s$, the
fluctuations outside the band $\omega,\omega',\omega'' \in \pm
[\omega_0-\delta \omega,\omega_0+\delta \omega]$ can be neglected
and within this band,
\begin{equation}
\begin{split}
R_I(\omega,\omega',\omega'')=& \frac{Z_s^3}{(Z_s+Z_p)^3}
\Bigg[\left(\frac{R_1}{2} + \frac{3D_1}{2}
\frac{Z_p}{(Z_s+Z_p)^2}\left(2k_B T_0 - Z_s S_0\right)
-\frac{3Z_s}{4} \sum_{m\neq 0, n \pm m =1} D_n
S_m\right)\delta(\omega_s \pm \omega_0)
\\&+ \left(\frac{R_3}{2} + \frac{3D_3}{2} \frac{Z_p}{(Z_s+Z_p)^2} \left(2k_B T_0 - Z_s S_0
\right) - \frac{3Z_s}{4} \sum_{m\neq 0, n \pm m = 3}D_n
S_m\right)\delta(\omega_s \pm 3\omega_0) \Bigg],
\end{split}
\end{equation}
Other Fourier components of $R$ and $D$ vanish due to the
filtering provided by $Z_e$.

Another important observable is the third moment of voltage
fluctuations over the impedance $Z_0$. This is given by
\begin{equation}
\begin{split}
R_V(\omega,\omega',\omega'') = &\frac{Z_s^3
Z_0(\omega)Z_0(\omega')Z_0(\omega'')}{Z(\omega)Z(\omega')Z(\omega'')}
\Bigg[-R_I^s + Z_s k_B T_0
\left(\frac{1}{Z(\omega)}+\frac{1}{Z(\omega')}+\frac{1}{Z(\omega'')}\right)
\sum_{n,\alpha=\pm 1} D_n \delta(\omega_s + n\alpha\omega_0) \\
&+ \frac{Z_s}{4} \sum_{\overset{n,m}{\alpha,\beta=\pm 1}} D_n S_m
\left(\frac{Z_0(\omega+m\beta  \omega_0)}{Z(\omega+m\beta
\omega_0)}+\frac{Z_0(\omega'+m\beta  \omega_0)}{Z(\omega'+m\beta
\omega_0)}+\frac{Z_0(\omega''+m\beta  \omega_0)}{Z(\omega''+m\beta
\omega_0)}\right) \delta(\omega_s+(n \alpha + m
\beta)\omega_0)\Bigg].
\end{split}
\end{equation}
Now the pass-band may be defined through $Z_p \approx Z_s$ and the
stop-band by $Z_e \ll Z_s$. Assume for simplicity that $Z_p=Z_s$.
Then, if all the frequencies are within the pass-band, one obtains
\begin{equation}
R_V(\omega,\omega',\omega'')=\frac{Z_s^3}{8}\left[-\frac{R_1}{2}\delta(\omega_s
\pm \omega_0) - \frac{R_3}{2}\delta(\omega_s \pm 3\omega_0)+
\frac{3}{4}(Z_s S_0 + 2k_B T_0)\left(D_1 \delta(\omega_s \pm
\omega_0) + D_3 \delta(\omega_s \pm 3\omega_0)\right)\right].
\end{equation}
\end{widetext}
In the stop-band, $R_V$ is negligible. \cite{matchingnote}

As an example, consider a linear system with Fano factors $F_2$
and $F_3$ (for the second and third moments) in an
impedance-matched environment $Z_p=Z_s$ and driven with $V(t)=V_3
\cos(3\omega_0 t)$, $eV_3 \gg k_B T$. The third moment of voltage
fluctuations over the sample within the pass-band is
\begin{equation}
R_V=Z_s^3\left[-\frac{e^2 F_3 I_3}{16} + \frac{3}{4\pi^2} F_2^2
e^2 I_3 + \frac{3}{\pi} F_2 e \frac{k_B T_0}{Z_s}\right]
\delta(\omega_s \pm 3\omega_0).
\end{equation}
This equation illustrates how the measured third moment consists
of the "real signal" (dependent on $F_3$) and the parts coming
from the sample noise driven by its own fluctuations (second term)
and those in the environment (third term).

In their paper,\cite{reulet03} Reulet {\it et al.} showed that the
environmental effect can be accurately described by this theory
after including the finite propagation time $\tau$ in the coaxial
cable between the filter and the sample (see
Fig.~\ref{fig:environmenteffect}). Their analysis is analogous in
the case of slow driving, and indicates that a long cable can be
used to decrease the effective noise temperature $T_0$ of the
environment.

\subsection{Averaging time}
\label{subs:averaging} In Ref.~\onlinecite{reulet03}, the required
time for signal averaging was several hours for each plot.
Naturally, for any reasonable measurement, this time should not be
much longer. To get an estimate of the required measurement time
with the cyclostationary driving, we consider the AC statistics
scheme (the requirements are analogous for the first scheme, only
the prefactors of the expressions may slightly differ). There, one
obtains a single output value in a time determined by the data
integration time $t_{\rm meas}=2\pi/\omega_m$. Assume one has
measured $n$ values of current. The squared deviation between the
measured value of the unbiased estimator $k_3$ for the third
moment of the data and the true third moment can be estimated
through the variance of $k_3$.\cite{weissteinworldofmath} In the
limit $n \gg 1$ it is given by
\begin{equation}
{\rm var}(k_3)=\frac{1}{n}(9\mu_2^3-\mu_3^2-6\mu_2 \mu_4 + \mu_6),
\end{equation}
where $\mu_i$ are the $i$'th central moments of the true current
distribution. Now let us require that the relative error is
smaller than some percentage $p$ of $\mu_3$,
\begin{equation}
{\rm var}(k_3) < p^2 \mu_3^2.
\end{equation}
This leads to the requirement
\begin{equation}
n >
\frac{1}{p^2}\left(\frac{9\mu_2^3}{\mu_3^2}-1-6\frac{\mu_2\mu_4}{\mu_3^2}+\frac{\mu_6}{\mu_3^2}\right).
\label{eq:nestimate}
\end{equation}
From this example, it is thus clear that any additional noise from
the measuring setup increases the measurement time through the
increase of the even moments.

In the typical limit where the current $I$ through the sample
exceeds $e$ times the bandwidth $\omega_d$ (e.g., for $I=100$ nA,
$I/e=620$ GHz), it is safe to neglect the cumulants of higher
order than three, i.e., the measured data is almost Gaussian. In
this case the above requirement reduces to
\begin{equation}
n > \frac{6}{p^2}\frac{\mu_2^3}{\mu_3^2} = \frac{32\pi}{3 p^2
\omega_d}\frac{(2 S_0+S_2 \cos(2 \omega_0 t_0))^3}{(3 R_1
\cos(\omega_0 t_0)+R_3 \cos(3 \omega_0 t_0))^2},
\end{equation}
where the last form was obtained using
Eqs.~(\ref{eq:secondm},\ref{eq:thirdm}).

To get an idea of the required measuring times, consider a linear
system driven with the voltage $V_3 \cos(3\omega_0 t)$ in the
shot-noise limit $eV_3 \gg k_B T$, and in the case when the sample
noise dominates the second and third moment of the fluctuations
(see Subs.~\ref{subs:linearsyst}). In this case the required
number of points for $t_0=0$ is
\begin{equation}
n > \frac{2048}{3 \pi^2 p^2} \frac{F_2^3}{F_3^2} \frac{I_3}{e
\omega_d}.
\end{equation}
Using the values $I_3=100$ nA, $\omega_m=\omega_d=2\pi ({\rm 10
MHz})$, and $p=0.05$ for a tunnel junction ($F_2=F_3=1$), we
obtain $n \gtrsim 2.7 \cdot 10^{8}$ and $T=2\pi n/\omega_m \gtrsim
27$ s.

If the current is lowered, also the measuring time is lowered as
the "skewness" $\mu_3/\mu_2^{3/2}$ increases with decreasing
average. Of course, the estimate is valid only in the shot-noise
limit $eV\gg k_B T$ and when the sample noise dominates the
amplifier noise. Therefore, optimal signal is expected for the
minimum values of current with which the signal moments are still
determined by shot noise. Note that typically increasing the
impedance of the measured sample decreases the possible bandwidth
for impedance matching. However, also the shot-noise limit $eV \gg
k_B T$ is obtained with a lower current. As a result, the overall
averaging time for the third moment does not change much.

If the amplifier noise $S_I^0 = 2 k_B T_0/Z_0$ dominates the shot
noise, we get in the otherwise same case as above the requirement
\begin{equation}
n>\frac{2048 \pi}{3p^2}\left(\frac{k_B T_0}{e I_3 Z_0}\right)^3
\frac{I_3}{e \omega_d}.
\end{equation}
Now, as an example, for $T_0=1 K$, $I_3=1 \mu A$, $Z_0=50 \Omega$,
$\omega_m=\omega_d=2 \pi ({\rm 100 MHz})$ and $p=0.05$ the
required number of measurements for a tunnel junction would be $n
\gtrsim 4.4 \cdot 10^{10}$ and $T \gtrsim 440$ s.

A similar estimate for the second moment would yield a sample size
(in the limit $n \gg 1$)
\begin{equation}
n > \frac{1}{p^2 \mu_2^2}(\mu_4-\mu_2^2)
\end{equation}
and for the example considered above,
\begin{equation}
n > \frac{1}{p^2}\left(\frac{3\pi}{8}-1\right) \approx
\frac{0.17}{p^2}
\end{equation}
independent of the current or the bands (in the Gaussian limit
where the prefactor is slightly different, this is analogous to
the Dicke radiometer formula \cite{dicke}). Hence, for $p=0.01$,
we would need a few thousand samples, and the required measuring
time with MHz sampling would be of the order of milliseconds.



\section{Summary}
In this paper, we detail methods to measure the moments of current
fluctuations by employing a nonstationary driving signal. Such
driving overcomes the requirement for the wide-band measurements
and gives a way to confirm the measurement results through the
variation of the phase between the signal and the mixing. However,
cyclostationary driving makes rise to higher harmonics of the
signal, which have to be taken care of through a proper Fourier
analysis. In Sect.~\ref{sect:examples}, we discuss a few examples
of such an analysis.

Section \ref{sect:limitations} shows that the measurement of the
third moment always involves the effect caused by the measuring
setup. The averaging times discussed at the end of the same
section indicate that this type of "classical" measurements are
practical for the measurement of the third and perhaps still for
the fourth cumulant, but higher cumulants seem to be out of the
time constants set for any reasonable project. Therefore, schemes
where the detector is a mesoscopic system and placed near the
sample \cite{tobiska03,sonin04}, overcoming the low-bandwidth
restrictions, are preferable.

Let us outline a practical scheme for the detection of the third
moment/cumulant of current fluctuations with cyclostationary
driving. Assume one measures a sample with (possibly
voltage-dependent) impedance $Z_s$.
\begin{enumerate}
\item Construct the matching circuit that approximately matches
the input impedance of the amplifier to that of the sample, and
filters the signal from the narrow band $\omega \in \pm
[\omega_0-\delta \omega, \omega_0 + \delta \omega]$. Note that the
measurement outcome will depend on the details of the matching, as
discussed in Subs.~\ref{subs:enveffect}.

\item Drive the sample with voltage $V(t)=V_0 + V_3 \cos(3\omega_0
t)$ where either $V_0=0$ or $V_0 \gg V_3$ ("differential
measurement", see Subs.~\ref{subs:genericnonlinear}).

\item Either mix the signal twice with itself and once with
$f(t)=\cos(3\omega_0 (t+t_0))$ (Subs.~\ref{subs:acmixing}) or only
with $f(t)=\cos(\omega_0 (t+t_0))$
(Subs.~\ref{subs:acstatistics}). Average the signal (mixing
scheme) or its third moment (statistics scheme) for time exceeding
that indicated in Subs.~\ref{subs:averaging}. The outcome is given
by Eq.~(\ref{eq:mixingresult}) or Eq.~(\ref{eq:thirdm}), where
$R_3$ is taken from the $\cos(3\omega_0 t)$-component of the
Fourier series of $R(V(t))$ as in Subs.~\ref{subs:linearsyst},
\ref{subs:finitetemp}.
\end{enumerate}

\section{Acknowledgements}
We thank Pertti Hakonen, Jukka Pekola, and Pauli Virtanen for
clarifying discussions.

\appendix

\section{Moments in the AC statistics scheme} \label{sect:secondschemeapp}
The contribution of the $l$'th harmonic of the $n$'th moment,
$M_n^l$, to the $n$'th measured moment in the AC statistics scheme
can be written as
\begin{widetext}
\begin{equation}
\langle (\delta Y)^n \rangle_l = \frac{M_n^l}{(2\pi)^{n}} \int
dt_1 \dots dt_n H_{LP}^m(t_1) \dots H_{LP}^m(t_n) f(t_1) \dots
f(t_n) \int dt'  H_{BP}(t_1-t') \dots H_{BP}(t_n-t')
\cos(l\omega_0 t').
\end{equation}
\end{widetext}
The convolution between the $H_{BP}$-functions filters out
most of the harmonics $l$. It only leaves $l=1,3,\dots,n$ in the
case when $n$ is odd and $l=0,2,\dots,n$ when $n$ is even.

Now, use the fact that $H_{BP}(t)=2\cos(\omega_0 t)
H_{LP}^\delta(t)$, where $H_{LP}^\delta$ is a low-pass filter with
band $\omega \in [-\delta \omega,\delta \omega]$. One obtains
\begin{widetext}
\begin{equation}
\langle (\delta Y)^n\rangle_l=M_n^l \left(\frac{1}{\pi}\right)^{n}
\int dt_1 \dots dt_n dt' H_{LP}^m(t_1) \dots H_{LP}^m(t_n)
H_{LP}^\delta(t_1-t') \dots H_{LP}^\delta(t_n-t')
G(t_1,\dots,t_n,t'), \label{eq:momint}
\end{equation}
where
\begin{equation}
G(t_1,\dots,t_n,t')=f(t_1)\dots f(t_n) \cos(\omega_0(t_1-t'))
\dots \cos(\omega_0(t_n-t')) \cos(l\omega_0 t').
\end{equation}
\end{widetext}
If $\omega_m,\delta \omega < \omega_0/n$, only the
time-independent part of $G(t_1,\dots,t_n,t')$ survives the
remaining filtering. This can be computed from
\begin{equation}
G_{DC}^{n,l} \equiv \left(\frac{\omega_0}{2\pi}\right)^{n+1}
\int_0^{2\pi/\omega_0} dt_1\dots dt_n dt' G(t_1,\dots,t_n,t').
\end{equation}
However, $G_{DC}^{n,l}$ may still depend on the relative phase
$t_0$ between the driving and the mixed function
$f(t)=\cos(\omega_0(t+t_0))$. Equivalently to above,
$G_{DC}^{n,l}$ is only finite if $l$ is one of the values that
survives band-pass filtering. Moreover, it is straightforward to
show that $G_{DC}^{n,l}(t_0) = g_n^l \cos(l\omega_0 t_0)/2^{n+1}$.
The number $g_n^l$ for some of the lowest (nontrivial) values of
$n,l$ is tabulated in Table \ref{tab:gdctable}.

\begin{table}[h]
\begin{tabular}{c | c c c c c c c c c c c c c c c}
\hline $n$ & 1 & 2 & 2 & 3 & 3 & 4 & 4 & 4 & 5 & 5 & 5 & 6 & 6 & 6
& 6\\\hline $l$ & 1 & 0 & 2 & 1 & 3 & 0 & 2 & 4 & 1 & 3 & 5 & 0 &
2 & 4 & 6\\\hline $g_n^l$ & 1 & 1 & $\frac{1}{2}$ & $\frac{3}{4}$
& $\frac{1}{4}$ & $\frac{3}{4}$ & $\frac{1}{2}$ & $\frac{1}{8}$ &
$\frac{5}{8}$ & $\frac{5}{16}$ & $\frac{1}{16}$ & $\frac{5}{8}$ &
$\frac{15}{32}$ & $\frac{3}{16}$ & $\frac{1}{32}$\\\hline
\end{tabular}
%
\caption{Values of $g_n^l$ for different moments $n$ and different
harmonics $l$.} \label{tab:gdctable}
\end{table}

The remaining part of the integral in Eq.~(\ref{eq:momint}) is now
a combination of low-pass filters:
\begin{widetext}
\begin{equation}
\langle (\delta Y)^n\rangle_l = M_n^l g_n^l \cos(l\omega_0 t_0)
\frac{1}{2}\left(\frac{1}{2\pi}\right)^{n} \int dt_1 dt_2 \dots
dt_n dt' H_{LP}^m(t_1) \dots H_{LP}^m(t_n) H_{LP}^\delta(t_1-t')
\dots H_{LP}^\delta(t_n-t').
\end{equation}
\end{widetext}
Singling out one of the $t_i$ integrals, it is straightforward to
show that
\begin{equation}
\begin{split}
\int dt_i H_{LP}^m(t_i) H_{LP}^\delta(t_i-t') &= \sqrt{2\pi} {\cal
F}^{-1}_{t'}\left[\tilde{H}_{LP}^m(\omega)
\tilde{H}_{LP}^\delta(\omega)\right]
\\&= \sqrt{2\pi} H_{LP}^d(t'),
\end{split}
\end{equation}
where $H_{LP}^d(t')$ is a low-pass filter with bandwidth $\omega_d
= \min(\omega_m,\delta \omega)$ ($\omega \in
[-\omega_d,\omega_d]$). Thus we get
\begin{equation}
\langle (\delta Y)^n\rangle_l = \frac{1}{2}
\left(\frac{1}{2\pi}\right)^{n/2} M_n^l g_n^l \cos(l\omega_0 t_0)
{\cal H}_n(\omega_d).
\end{equation}
Here
\begin{equation}
\begin{split}
&{\cal H}_n(\omega_d) = \int dt' H_{LP}^d(t')^n =
\frac{1}{(2\pi)^{n/2-1}}\times \\&\int d\omega_1 \dots
d\omega_{n-1} H_{LP}^d(\omega_1) \dots H_{LP}^d(\omega_{n-1})
H_{LP}^d(\omega_1+\dots \omega_{n-1}) \\&\equiv \frac{\alpha_n
\omega_d^{n-1}}{(2\pi)^{n/2-1}}.
\end{split}
\end{equation}
and $\alpha_n$ is a number whose values for lowest $n$ are
\begin{equation}
\begin{split}
\alpha_1 &= 1 \quad \quad \quad \quad \quad \quad \quad \quad \alpha_2 = 2\\
\alpha_3 &= 3 \quad \quad \quad \quad \quad \quad \quad \quad \alpha_4 = 16/3\\
\alpha_5 &= 115/12 \quad \quad \quad \quad \quad \alpha_6 = 88/5\\
\alpha_7 &= 17407/360 \quad \quad \quad \quad \alpha_8 = 53752/315\\
\alpha_9 &= 18063361/40320 \quad \alpha_{10} = 1440.
\end{split}
\end{equation}
Finally the contribution of the $l$'th harmonic of the $n$'th
moment on the $n$'th moment of the measured quantity is given by
\begin{equation}
\langle (\delta Y)^n \rangle = \frac{1}{2^n\pi^{n-1}} g_n^l
\alpha_n \omega_d^{n-1} M_n^l \cos(l\omega_0 t_0).
\end{equation}
The lowest six moments are listed in
Eqs.~(\ref{eq:firstm}-\ref{eq:sixthm}).


\begin{thebibliography}{99}
\bibitem{numbertwo} Throughout, we omit the number two in the definition of
the correlation functions for the second moment, used by many
authors to account for the positive and negative frequencies.
\bibitem{matchingnote} Note that in a typical measurement, the
voltage fluctuations are read only through the dissipative part of
the environment, whereas impedance matching requires transforming
this resistance $R_d$ to a value $Z_p$ near the sample impedance.
This is typically done with an LC circuit. Hence, the resulting
voltage fluctuations in the amplifier have to be scaled by the
ratio between $R_d$ and $Z_p$.
\bibitem{optimumnote} The optimal signal-to-noise ratio is
typically not obtained with this choice, but this choice is made
here for simplified notation.
\end{thebibliography}
\end{document}